\documentclass[letterpaper,10pt,reqno]{amsart}
\usepackage[utf8x]{inputenc}
\usepackage{amsmath, amssymb, amsfonts}
\usepackage[foot]{amsaddr}
\usepackage{amsthm}
\newtheorem{conjecture}{Conjecture}[section]
\newtheorem{proposition}{Proposition}[section]
\newtheorem{lemma}{Lemma}[section]
\newtheorem{theorem}{Theorem}[section]
\newtheorem{corollary}{Corollary}[section]
\usepackage{enumerate}
\usepackage{titlesec}
\usepackage[margin=2.5cm]{geometry}
\numberwithin{equation}{section}
\titleformat{\subsection}
  {\centering}{\thesubsection}{1em}{}
\titleformat{\section}
  {\normalfont\scshape\centering}{\thesection.}{1em}{}

\title[An Asymptotic Expansion and Recursive Inequalities for the Monomer-Dimer Problem]{An Asymptotic Expansion and Recursive Inequalities\\for the Monomer-Dimer Problem}
\author{Paul Federbush$^*$}
\address{$^*$Department of Mathematics, University of Michigan, Ann Arbor, MI 48109-1043, \textit{email:} pfed@umich.edu}
\author{Shmuel Friedland$^{\dagger}$}
\address{$^{\dagger}$Department of Mathematics, Statistics and Computer Science, University of Illinois at Chicago, Chicago, IL 60607-7045, \textit{email:} friedlan@uic.edu }

\date{February, 2011}

\begin{document}

\maketitle

\begin{abstract}
Let $\lambda_d(p)$ be the $p$ monomer-dimer entropy on the $d$-dimensional integer lattice ${\mathbb Z}^d$,
where $p\in[0,1]$ is the dimer density.
We give upper and lower bounds for $\lambda_d(p)$ in terms of expressions involving $\lambda_{d-1}(q)$.
The upper bound is based on a conjecture claiming that the $p$ monomer-dimer entropy of an infinite
subset of ${\mathbb Z}^d$ is bounded above by $\lambda_d(p)$.  We compute the first three terms in the formal
asymptotic expansion of $\lambda_d(p)$ in powers of $\frac{1}{d}$.  We prove that the lower asymptotic matching conjecture is satisfied for $\lambda_d(p)$.
\end{abstract}
\textbf{Keywords and phrases:}  monomer-dimer entropies: asymptotic expansion, recursive inequalities and
a conjecture.

\section{Introduction}
\label{Intro}
The first aim of this paper is to discuss an asymptotic expansion of the monomer-dimer $p$-entropy, denoted by $\lambda_d \left( p \right)$, where $p \in \left[ 0,1 \right] $ is the density of dimers, on the integer $d$-dimensional lattice $\mathbb{Z}^d$.  The study of the existence of this entropy, some of its properties and its estimates, was initiated in a series of papers by Hammersley and his collaborators \cite{HammExist, HammImprove, HammCalc, HammMenon}.  It was shown by Minc \cite{Minc} that the $d$-dimensional dimer entropy $\lambda_d \left( 1 \right)$ satisfies
\begin{align}
\frac{1}{2} \log \left( 2d \right) - \frac{1}{2} \le \lambda_d \left( 1 \right) \le \frac{\log \left( 2d \right)!}{4d} \le \frac{1}{2} \log \left( 2d \right) - \frac{1}{2} + \frac{\log 2 \pi \left( 2d \right)}{4d} + \frac{1}{48d^2}.
\end{align}
The lower bound is implied by the proof of the van der Waerden conjecture \cite{Egorichev, Falikman}, or its \textit{weak} form \cite{Friedland}.  The first upper bound follows from the Bregman ineqality \cite{Bregman}.  The last upper bound follows from a sharp form of the Stirling formula for $\left( 2d \right)!$ \cite[p 52]{Feller}.  In a series of papers \cite{FedHid} -- \cite{FedTerm}, the first author studied a possible asymptotic expansion of
\begin{align}
\label{possexp}
\lambda_d \left( 1 \right) \sim \frac{1}{2} \log \left( 2d \right) - \frac{1}{2} + \sum_{k=1}^{\infty} \frac{c_k}{d^k}.
\end{align}
This is derived assuming that an argument employing a formal cluster expansion could be made rigorous.  He was able to compute the values of $c_1, c_2, c_3$, equal to $\frac{1}{8}, \frac{5}{96}, \frac{5}{64}$ respectively.  In this paper we show that the methods in \cite{FedHid} -- \cite{FedTerm} can be also applied to the asymptotic expansion of $\lambda_d \left( p\right)$ for any $p \in \left[ 0,1 \right]$.  The lower and the upper bounds for $\lambda_d \left( p \right)$ given in \cite[(5.3)]{FriedPeledComp} and \cite[(7.5)]{Friedlandetal} respectively yield
\begin{align}
\frac{1}{2} \left( p \log 2d - p \log p - 2 \left( 1 - p \right) \log \left( 1 - p \right) - p \right) \le \lambda_d \left( p \right) & \le \\
\frac{1}{2} \left( p \frac{\log \left( 2d \right)!}{2d} - p \log p - 2 \left( 1 - p \right) \log \left( 1 - p \right) \right) & \le \\
\frac{1}{2} \left( p \log 2d - p \log p - 2 \left( 1 - p \right) \log \left( 1 - p \right) - p \right) + p \left( \frac{\log 2 \pi \left( 2d \right)}{4d} + \frac{1}{48d^2} \right). \notag
\end{align}
Assuming valid the same formal procedure as in \cite{FedHid} or \cite{FedTerm}, we show that
\begin{align}
\label{lambdaseries}
\lambda_d \left( p \right) \sim \frac{1}{2} \left( p \log 2d - p \log p - 2 \left( 1 - p \right) \log \left( 1 - p \right) - p \right) + \sum_{k=1}^{\infty} \frac{c_k \left( p \right)}{d^k},
\end{align}
where each $c_k \left( p \right)$ is a polynomial in $p$.
This polynomial, $c_k(p)$, is built of powers $p^s$ with  $k < s \le 2k$ ( as follows from the form of equation \eqref{expandJs} below ).

We compute the polynomials $c_1 \left( p \right), c_2 \left( p \right), c_3 \left( p \right)$,
\begin{align}
c_1 \left( p \right) &= \tfrac{1}{8} p^2 \label{c1} \\
c_2 \left( p \right) &= \left( 2 p^3 + 3 p^4 \right) / 96 \label{c2} \\
c_3 \left( p \right) &= \left( - 5 p^4 + 12 p^5 + 8 p^6 \right) / 192. \label{c3}
\end{align}
While studying $\lambda_d \left( p \right)$ we made a heuristic observation
\begin{align}
\label{heurobs}
\lambda_d \left( p \right) \cong \lambda_{d-1} \left( p \left( 1 - \tfrac{1}{d} \right) \right) + \left( 1 - p \left( 1 - \tfrac{1}{d} \right) \right) \log \left( 1 - p \left( 1 - \tfrac{1}{d} \right) \right) - \tfrac{p}{2d} \log \left( \tfrac{p}{2d} \right) \notag \\
+ \left( 1 - \tfrac{p}{2d} \right) \log \left( 1 - \tfrac{p}{2d} \right) - \left( 1 - p \right) \log \left( 1 - p \right),
\end{align}
for large $d$.  We also substituted the ansatz
\begin{align}
\label{ansatz}
\lambda_d \left( p \right) = \frac{1}{2} p \log \left( 2d \right) - \frac{p}{2} \log p - \left( 1 - p \right) \log \left( 1 - p \right) - \frac{p}{2} + \frac{ap^2}{d} + \frac{bp^4 + cp^3}{96d^2}
\end{align}
into both sides of \eqref{heurobs}, and found the value of the right side of the equation minus the left side of the equation is equal to
\begin{align}
\label{rightminusleft}
- \frac{1}{8} \left( -1 + 8a \right) \frac{p^2}{d^2} - \frac{1}{96} \left( -2 + 2bp + c \right) \frac{p^3}{d^3} + \mathcal{O} \left( \frac{1}{d^4} \right)
\end{align}
The correct values $a = \frac{1}{8}$ and $c=2$ are determined to achieve accuracy up to order $\frac{p^4}{d^3}$ (other choices would have terms with lower powers of $p$ or $\frac{1}{d}$).  We view this result as both a measure of how good an approximation \eqref{heurobs} is, and a verification that the series \eqref{lambdaseries} is correct (to a certain order).

We show in this paper that \eqref{heurobs} is actually a recursive inequality
\begin{align}
\label{recur}
\lambda_d \left( p \right) \ge \lambda_{d-1} \left( p \left( 1 - \tfrac{1}{d} \right) \right) & + \left( 1 - p \left( 1 - \tfrac{1}{d} \right) \right) \log \left( 1 - p \left( 1 - \tfrac{1}{d} \right) \right) - \tfrac{p}{2d} \log \left( \tfrac{p}{2d} \right) \notag \\
& + \left( 1 - \tfrac{p}{2d} \right) \log \left( 1 - \tfrac{p}{2d} \right) - \left( 1 - p \right) \log \left( 1 - p \right).
\end{align}
We also show the validity of some upper bounds on $\lambda_d \left( p \right)$ in terms of the function $\lambda_{d-1} \left( p \right)$:
\begin{align}
\label{upperbound}
& \lambda_d \left( p \right) \le \notag \\
& \max_{u \in \left[0, \frac{p}{d} \right]} \left\{ \left( 1 - \tfrac{u}{2} \right) \log \left( 1 - \tfrac{u}{2} \right) - \tfrac{u}{2} \log \tfrac{u}{2} - \left( 1 - u \right) \log \left( 1 - u \right) + \left( 1 - u \right) \lambda_{d-1} \left( \frac{p-u}{1-u} \right) \right\}
\end{align}
and
\begin{align}
\label{betterbound}
& \lambda_d \left( p \right) \le \notag \\
& \max_{u \in \left[ 0,\frac{p}{d} \right]}\left\{ \left( 1-p+\tfrac{u}{2} \right) \log \left( 1-p+\tfrac{u}{2} \right)- \tfrac{u}{2} \log \tfrac{u}{2} - \left( 1-p \right) \log \left( 1-p \right)   + \left( 1 - \tfrac{u}{2} \right) \lambda_{d-1} \left( \frac{p-u}{1-\frac{u}{2}} \right) \right\}
\end{align}
if the following conjecture holds.  
\begin{conjecture}
Let $\mathcal{S} = \left\{ S_i, i \in \mathbb{N} \right\}$ be an increasing sequence of finite sets in $\mathbb{Z}^d$.  We can define its $p$ monomer-dimer entropy $\lambda_{\mathcal{S}} \left( p \right)$ similar to the definition of $\lambda_d \left( p \right)$.  We conjecture that $\lambda_{\mathcal{S}} \left( p \right) \le \lambda_d \left( p \right)$.
\end{conjecture}
This conjecture will be precisely stated in Section \ref{Prelim} which contains some preliminary formalism.  The lower bound is studied in Section \ref{Lower}.  Section \ref{Upper} deals with the upper bounds, and goes a long way towards proving that the bound of \eqref{betterbound} is always better than the bound of \eqref{upperbound}, as we believe.  The asymptotic expansion is developed in Section \ref{Asymptotic}.  Section \ref{Conclusions} studies some implications of our inequalities, in particular proving the LAMC for the monomer-dimer problem on a rectangular lattice, see \cite{Friedlandetal}.  Section \ref{numtests} studies numerical tests of our bounds and expansion for $d=1,2,3$.  Although our expansion was developed as an asymptotic expansion in $1 / d$ it has surprising validity for small $d$.

\section{Preliminary Results}
\label{Prelim}
Let $\mathbb{Z}^d$ be the $d$-dimensional  integer lattice in $\mathbb{R}^d$.  Two points $\mathbf{x} = \left( x_1 , \dots , x_d \right), \mathbf{y} = \left( y_1 , \dots , y_d \right) \in \mathbb{Z}^d$ are called neighbors if $\| \mathbf{x} - \mathbf{y} \|_1 = \sum_{i=1}^{d} \left| x_i - y_i \right| = 1$.  Let $G = \left( \mathbb{Z}^d, E_d \right)$ be the infinite graph whose vertices are $\mathbb{Z}^d$, and whose edges, $E_d$, are the set of pairs $\left( \mathbf{x} , \mathbf{y} \right)$ where $\mathbf{x}, \mathbf{y}$ are neighbors.

Note that the degree of each vertex is $2d$.  Set $\left[ m \right] = \left\{ 1, \dots , m \right\}$ for a positive integer $m$.  Then $\left[ \mathbf{m} \right] = \left[m_1\right] \times \left[m_2\right] \times \dots \times \left[m_d\right] $ is a box containing $\# \left[ \mathbf{m} \right] = m_1 m_2 \dots m_d$ points in $\mathbb{Z}^d$, where $\mathbf{m} = \left( m_1, \dots , m_d \right) \in \mathbb{N}^d$ has positive integer coordinates.  We view the finite lattice $\left[ \mathbf{m} \right]$ as a graph $G \left( \mathbf{m} \right) = \left( \left[ m \right] , E \left( \left[ \mathbf{m} \right] \right) \right)$ where $\mathbf{x}, \mathbf{y} \in \left[ \mathbf{m} \right]$ are neighbors, i.e. $\left( \mathbf{x}, \mathbf{y} \right) \in E \left( \mathbf{m} \right)$ if and only if $\| \mathbf{x} - \mathbf{y} \| = 1$.  Let $T \left( \mathbf{m} \right)$ be a $d$-dimensional torus on the finite lattice $\left[ \mathbf{m} \right]$, where $T \left( \mathbf{m} \right)$ is obtained from the infinite graph $G = \left( \mathbb{Z}^d , E_d \right)$ by considering the quotient $\mathbb{Z}^d / \left( \left( m_1 \mathbb{Z} \right) \times \dots \times \left( m_d \mathbb{Z} \right)\right) $.  So $T \left( \mathbf{m} \right)$ is a $2d$ regular graph, and $G \left( \mathbf{m} \right)$ is a subgraph of $T \left( \mathbf{m} \right)$.

A monomer-dimer tiling of $\left[ \mathbf{m} \right]$, also called a configuration $\phi$ in $\left[ \mathbf{m} \right]$, consists of $\ell$ nonoverlapping dimers placed on the neighboring lattice points in $\left[ \mathbf{m} \right]$.  Other lattice points in $\left[ \mathbf{m} \right]$ are viewed as covered by monomers.  Denote by $C \left( \left[ \mathbf{m} \right] , \ell \right)$ the space of monomer-dimer tilings $\phi$ of $\left[ \mathbf{m} \right]$ with $\ell$ dimers.  Denote by $C_{\mathrm{per}} \left( \left[ \mathbf{m} \right] , \ell \right)$ all monomer-dimer tilings of $T \left( \mathbf{m} \right)$ with $\ell$ dimers.  Clearly, $C \left( \left[ \mathbf{m} \right] , \ell \right) \subseteq C_{per} \left( \left[ \mathbf{m} \right], \ell \right)$.  Let $\# C \left( \left[ \mathbf{m} \right] , \ell \right) , \# C_{\mathrm{per}} \left( \left[ \mathbf{m} \right] , \ell \right)$ be the cardinalities of $C \left( \left[ \mathbf{m} \right] , \ell \right), C_{\mathrm{per}} \left( \left[ \mathbf{m} \right] , \ell \right)$ respectively.  It was shown by Hammersley \cite{HammExist}
\begin{align}
\label{Hammersley}
\lambda_d \left( p \right) = \lim_{\mathbf{m}_i \to \infty} \frac{\log C \left( \left[ \mathbf{m}_i \right] , \ell_i \right)}{\# \left[ \mathbf{m}_i \right]} = \lim_{\mathbf{m}_i \to \infty} \frac{\log C_{per} \left( \left[ \mathbf{m}_i \right] , \ell_i \right)}{\# \left[ \mathbf{m}_i \right]}
\end{align}
where $\mathbf{m}_i = \left( m_{i,1}, \dots , m_{i,d} \right) \in \mathbb{N}^d$, $\lim_{i \to \infty} m_{i,k} = \infty, k \in \left[ d \right]$, $\lim \frac{2 \ell_i}{\# \left[ \mathbf{m}_i \right]} = p \in \left[ 0,1 \right]$.
This equality does not depend on a particular choice of sequences $\mathbf{m}_i, \ell_i, i \in \mathbb{N}$ which satisfy the stated conditions.  For more detailed analysis of the above limit and upper and lower bounds of $\lambda_d \left( p \right)$ see \cite{FriedPeledComp}.

We now discuss the notion of $p$ monomer-dimer entropy $\lambda_{\mathcal{S}} \left( p \right)$ of a sequence $\mathcal{S} = \left\{ S_i, i \in \mathbb{N} \right\}$ of finite sets in $\mathbb{Z}^d$.  Assume that we have a sequence of increasing boxes $\left[ \mathbf{m}_i \right], i \in \mathbb{N}$, where $\mathbf{m}_i = \left( m_{1,i}, \dots m_{d,i} \right), i \in \mathbb{N}$ and for each $k \in \left[ d \right]$ the sequence $m_{i,k}, i \in \mathbb{N}$ increases to infinity.  Without loss of generality we may assume that $S_i \subseteq \left[ \mathbf{m}_i \right]$.  Then $S_i  = \left[ \mathbf{m}_i \right] \backslash U_i$, for a sequence of subsets $U_i \subseteq \left[ \mathbf{m}_i \right], i \in \mathbb{N}$.  Let $\mathcal{U} = \left\{ \left( \left[ \mathbf{m}_i \right], U_i \right) , i \in \mathbb{N} \right\}$.  It would be convenient to replace $\mathcal{S}$ by $\mathcal{U}$.  With $S_i$ we associate a subgraph $G_i$ of $T \left( \mathbf{m}_i \right)$ induced by the set of vertices $S_i = \left[ \mathbf{m}_i \right] \backslash U_i$.  (So $\mathbf{u}_i, \mathbf{v}_i \in \left[ \mathbf{m}_i \right] \backslash U_i$ are neighbors if and only if they are neighbors in $T \left( \mathbf{m}_i \right)$.)  Let $\mathcal{G} = \left\{ G_i , i \in \mathbb{N} \right\}$.  Then $\lambda_{\mathcal{S}} \left( p \right) = \lambda_{\mathcal{U}} \left( p \right)$ is defined as equal to $\lambda_{\mathcal{G}} \left( p \right)$ as in \cite[\S 2]{Friedlandetal}.  We now state this definition.

Let $\left[ \mathbf{m} \right] \subset \mathbb{N}^d, \mathbf{m} = \left( m_1, \dots, m_d \right) \in \mathbb{N}^d$ be a box in $\mathbb{Z}^d$.  Assume that $U \subseteq \left[ \mathbf{m} \right]$.  Denote by $\# \left( \left[ \mathbf{m} \right] \backslash U \right)$ the cardinality of the set $\left[ \mathbf{m} \right] \backslash U$.  Denote by $C \left( \left[ \mathbf{m} \right] \backslash U, \ell \right)$ all tilings of $\left[ \mathbf{m} \right] \backslash U$ with exactly $\ell \left( \le \frac{\# \left[ \mathbf{m} \right] \backslash U}{2} \right)$ dimers, when we view $\left[ \mathbf{m} \right]$ as the graph $G \left( \mathbf{m} \right)$.  (It is possible that $\# C \left( \left[ \mathbf{m} \right] \backslash U , \ell \right) = 0$).  Then
\begin{align}
& \lambda_{\mathcal{U}} \left( p \right) = \notag \\
& \limsup_{i \to \infty} \frac{\log \# C \left( \left[ \mathbf{m}_{i} \right] \backslash U_i , \ell_i \right)}{\# \left( \left[ \mathbf{m}_{i} \right] \backslash U_{i} \right)} \mathrm{ \ provided \ } \lim_{i \to \infty} \frac{2 \ell_i}{\# \left( \left[ \mathbf{m}_{i} \right] \backslash U_{i} \right)} = p.
\end{align}
Note that it is possible that $\lambda_{\mathcal{U}} \left( p \right) = - \infty$ for some $p \in \left[ 0,1 \right]$.  We now state a conjecture, which seems to be very reasonable.

\begin{conjecture}
\label{estim}
Let $d \in \mathbb{N}, \left[ \mathbf{m}_i \right], \mathbf{m}_i = \left( m_{1,i}, \dots , m_{d,i} \right), i \in \mathbb{N}$ be a sequence of increasing boxes $\left[ \mathbf{m}_i \right], i \in \mathbb{N},$ where $\lim_{i \to \infty} m_{k,i} = \infty, k = 1, \dots , d.$  Assume that $U_i \subseteq \left[ \mathbf{m}_i \right] , i \in \mathbb{N}$ and denote $\mathcal{U} = \left\{ \left( \left[ \mathbf{m}_i \right], U_i \right), i \in \mathbb{N} \right\}$.  Then
\begin{align}
\lambda_{\mathcal{U}} \left( p \right) \le \lambda_d \left( p \right) \mathrm{\ for \ each \ } p \in \left[ 0,1 \right].
\end{align}
\end{conjecture}

\begin{proposition}
Conjecture \ref{estim} holds for $d = 1$.
\end{proposition}

\textbf{Proof}.  Clearly $\left[ m \right] \backslash U$ consists of at most $\# U + 1$ disjoint paths, $P_1, \dots, P_k$.  Let $m' = \# \left( \left[ m \right] \backslash U \right)$ and view $G \left( \left[ m' \right] \right)$ obtained from $P_1, \dots, P_k$ by connecting by an edge the last vertex of $P_i$ and the first vertex of $P_{i+1}$ for $i = 1, \dots, k-1$.  Hence, each monomer-dimer tiling of $\left[ m \right] \backslash U$ is a monomer-dimer tiling of $\left[ m' \right]$. \qed

We assume that $d \ge 2$ and follow the discussion in \cite{FriedPeledComp}.  Let $\mathbf{m} = \left( m_1, \dots, m_d \right)$ and denote $\mathbf{m'} = \left(m_1, \dots, m_{d-1} \right), N' = \# \left[ \mathbf{m'} \right], m_d = n$.  Then the $\left( d-1 \right)$-dimensional box of the form $\left( \left[ \mathbf{m'} \right],j \right)$ is called the $j^{\mathrm{th}}$ layer, for $j=1, \dots, n$.  So $n=1$ corresponds to the $\left( d-1 \right)$-dimensional case.  Assume that $n \ge 2$ unless stated otherwise.  We view the adjacency graph $G \left( \mathbf{m} \right)$ as composed of $n$ layers of $G \left( \mathbf{m'} \right) $ plus the edges between the adjacent layers $\left( \left[ \mathbf{m'} \right],j \right)$ and $\left( \left[ \mathbf{m'} \right],j+1 \right)$ for $j=1, \dots , n-1$.  Let us consider one tiling of $\left[ \mathbf{m} \right]$ with $\ell$ dimers: $\phi \in C \left( \left[ m \right], \ell \right)$.  Each layer $\left( \left[ m' \right], j \right)$ has $h_i$ \textit{horizontal, (transversal)}, dimers for $j=1, \dots , n$.  Let $v_j$ be the number of \textit{vertical} dimers connecting the levels $\left( \left[ \mathbf{m'} \right], j \right)$ and $\left( \left[ \mathbf{m'} \right], j+1 \right)$ for $j = 1, \dots , n-1$.  Let us assume that $v_0 = v_n = 0$.  We have the following conditions for $n \ge 2$.
\begin{align}
\label{horver}
\sum_{j=1}^n \left( h_j + v_j \right) = \ell, \quad v_{j-1} + v_j + 2h_j \le N' \quad \mathrm{for} \quad i=1, \dots , n.
\end{align}
More precisely, the location of $v_j$ vertical dimers connecting the level $j$ and $j+1$ are $\left( V_j ,j \right)$ for a corresponding subset $V_j \subset \left[ \mathbf{m'} \right]$ for $j=1, \dots , n-1$.  So $v_j = \# V_j$.  We assume that $V_0 = V_n = \varnothing$.  Then $V_{j-1} \cap V_j = \varnothing$ for $j = 1, \dots, n$.  Denote $\mathbf{h} = \left( h_1 , \dots, h_n \right)$, $\mathbf{v} = \left( v_1 , \dots, v_{n-1} \right)$, $v_0 = v_n = 0$.  Let $C \left( \left[ \mathbf{m} \right], \mathbf{h}, \mathbf{v} \right) \subset C \left( \left[ \mathbf{m} \right], \ell \right)$ be all configurations of tilings of $\left[ \mathbf{m} \right], \mathbf{m} = \left( m_1 , \dots , m_{d-1}, n \right)$ with $\ell$ tiles, where $\mathbf{h}$ and $\mathbf{v}$ are the distributions of horizontal and vertical tiles.  (It is possible that $C \left( \left[ \mathbf{m} \right], \mathbf{h}, \mathbf{v} \right) = \varnothing$.)  Denote by $\mathcal{C} \left( \mathbf{m}, \ell \right)$ the set of all possible $\left( \mathbf{h}, \mathbf{v} \right)$ satisfying \eqref{horver} and the corresponding above stated conditions.  So
\begin{align}
C \left( \left[ \mathbf{m} \right], \ell \right) &= \cup_{\left( \mathbf{h}, \mathbf{v} \right) \in \mathcal{C} \left( \mathbf{m}, \ell \right)} C \left( \left[ \mathbf{m} \right] , \mathbf{h}, \mathbf{v} \right), \notag \\
\# C \left( \left[ \mathbf{m} \right], \ell \right) &= \sum_{\left( \mathbf{h}, \mathbf{v} \right) \in \mathcal{C} \left( \mathbf{m} , \ell \right)} \# C \left( \left[ \mathbf{m} \right], \mathbf{h}, \mathbf{v} \right) \label{cardsum}.
\end{align}
We claim that
\begin{align}
\label{cardCbound}
\# \mathcal{C} \left( \mathbf{m} , \ell \right) \le \left( N' \right)^{2n}.
\end{align}
Clearly $h_i \in \left[ 0, \frac{N'}{2} \right], v_i \in \left[ 0 , N' \right]$.  Hence the number of choices of $h_i, v_i$ is at most $\left( N'+1 \right) \frac{N'}{2}$, which is not greater than $\left( N' \right)^2$.  So \eqref{cardCbound} is a big overestimate, since we ignored \eqref{horver}.

We now show that to compute $\lambda_d \left( p \right)$ we need only to consider one sequence $\mathbf{h}_i, \mathbf{v}_i, k \in \mathbb{N}$ of dimer distributions.

\begin{lemma}
\label{seqlemma}
 Let $d \ge 2$ be an integer.  Assume that $\mathbf{m}_i \in \mathbb{Z}^d, \ell_i, i \in \mathbb{N}$ satisfies the conditions for \eqref{Hammersley}.  Then there exists a sequence
 \begin{align}
 \mathbf{h}_i &= \left( h_{1,i}, \dots, h_{m_{i,d},i} \right) \in \mathbb{Z}^{m_{i,d}}_+, \quad \mathbf{v}_i = \left( v_{1,i}, \dots , v_{\left( m_{i,d}-1 \right),i} \right) \in \mathbb{Z}^{\left( m_{i,d}-1 \right)}, \notag \\
 v_{i,0} &= v_{i,m_{i,d}}=0, \quad \sum_{j=1}^{m_{i,d}} h_{j,i}+ v_{j,i} = \ell_i, \quad i \in \mathbb{N}. \label{seqcond}
\end{align}
such that
\begin{align}
\lim_{i \to \infty} \frac{\log \# C \left( \left[ \mathbf{m}_i \right] , \mathbf{h}_i , \mathbf{v}_i \right)}{\# \left[ \mathbf{m}_i \right]} = \lambda_d \left( p \right)
\end{align}
\end{lemma}

\textbf{Proof.} Let $\left( \tilde{\mathbf{h}}_i , \tilde{\mathbf{v}}_i \right) , i \in \mathbb{N} $ be an allowable sequence satisfying \eqref{seqcond}.  The equalities \eqref{Hammersley} and \eqref{cardsum} yield the inequality
\begin{align}
\label{limsupcardbound}
\limsup_{i \to \infty} \frac{\log \# C \left( \left[ \mathbf{m}_i \right],\tilde{ \mathbf{h}}_i,\tilde{ \mathbf{v}}_i \right)}{\# \left[ \mathbf{m}_i \right]} \le \lambda_d \left( p \right).
\end{align}
Choose
\begin{align}
\left( \mathbf{h}_i , \mathbf{v}_i \right) = \mathrm{argument\; of} \max \left\{ \# C \left( \left[ \mathbf{m}_i \right] , \tilde{\mathbf{h}}_i, \tilde{\mathbf{v}}_i \right), \left( \tilde{\mathbf{h}}_i, \tilde{\mathbf{v}}_i \right) \in \mathcal{C} \left( \mathbf{m}_i , \ell_i \right) \right\} , i \in \mathbb{N}
\end{align}
i.e. a value of $\left( \tilde{\mathbf{h}}_i , \tilde{\mathbf{v}}_i \right)$ for which maximum is achieved.  Combine \eqref{cardsum} and \eqref{cardCbound} to deduce that
\begin{align}
\label{cardineq}
\# C \left( \left[ \mathbf{m}_i \right] , \ell_i \right) \le \left( \# \left[ \mathbf{m'}_i \right] \right)^{2m_{i,d}} \# C \left( \left[ \mathbf{m}_i \right] , \mathbf{h}_i , \mathbf{v}_i \right) , i \in \mathbb{N}.
\end{align}
Observe next that
\begin{align}
\label{limcardratio}
\lim_{i \to \infty} \frac{\log \left( \# \left[ \mathbf{m}'_i \right] \right)^{2m_{i,d}}}{\# \left[ \mathbf{m}_i \right]} = \lim_{i \to \infty} \frac{2 \log \# \left[ \mathbf{m'}_i \right)}{\# \left[ \mathbf{m'}_i \right]} =0
\end{align}
\eqref{cardineq} and \eqref{limcardratio} yield
\begin{align}
\liminf_{i \to \infty} \frac{\log \# C \left( \left[ \mathbf{m}_i \right] , \tilde{h}_i, \tilde{v}_i \right)}{\# \left[ \mathbf{m}_i \right]} \ge \lambda_d \left( p \right).
\end{align}
Combine this inequality with \eqref{limsupcardbound} to deduce the lemma.

\section{A Lower Estimate of $\lambda_d \left( p \right)$}
\label{Lower}
The main result of this section is:
\begin{theorem}
\label{mainestimate}
Let $d \ge 2$ be an integer and $p \in \left[ 0,1 \right]$.  Then
\begin{align}
\label{lambdalower}
\lambda_d \left( p \right) \ge \left( \lambda_{d-1} \left( q \right) + \left( 1-q \right) \log \left( 1-q \right) - \left( \frac{p-q}{2} \right) \log \left( \frac{p-q}{2} \right) + \right. \\
\left. \left( 1 - \frac{p-q}{2} \right) \log \left( 1 - \frac{p-q}{2} \right) - \left( 1-p \right) \log \left( 1-p \right) \right) \quad \mathrm{for \ each \ } q \in \left[ 0,p \right]. \notag
\end{align}
In particular \eqref{recur} holds.
\end{theorem}

We prove the theorem using a number of known results and straightforward lemmas.  Recall that $\lambda_d \left( p \right) \ge \lambda_{d-1} \left( p \right)$.  Hence to show \eqref{lambdalower} we need to consider the case $q<p$.  Next we recall the inequality
\begin{align}
\label{lambdalogineq}
\lambda_k \left( \frac{2 \ell}{\# \left[ \mathbf{m} \right]} \right) \ge \frac{\log C \left( \left[ \mathbf{m} \right], \ell \right)}{\# \left[ \mathbf{m} \right]}, k \in \mathbb{N}.
\end{align}
This follows by considering boxes $\left[ j \mathbf{m} \right]$ tiled with $j^d \ell$ dimers for $j = 1, \dots, $ \cite{HammExist, FriedPeledComp}.

Fix a rational number $q \in \left( 0,1 \right)$ and $\epsilon > 0$.  Then there exists $\mathbf{m'} \in \mathbb{N}^{d-1}, \ell' \in \mathbb{N}$ such that
\begin{align}
\label{mprimelambda}
q := \frac{2 \ell'}{\# \left[ \mathbf{m'} \right]}, \quad \lambda'_{d-1} \left( q \right) := \frac{\log C \left( \left[ \mathbf{m'} \right] , \ell' \right)}{\# \left[ \mathbf{m'} \right] } > \lambda_{d-1} \left( q \right) - \epsilon.
\end{align}
See \cite{HammExist}.  Let
\begin{align}
p = \frac{2L}{\# \left[ \mathbf{m'} \right]} \quad \mathrm{where \ } 2 \ell' < 2L \le \# \left[ \mathbf{m'} \right].
\end{align}
We will show for these choices of $p$ and $q$ the following inequality holds.
\begin{align}
\label{pqchoiceineq}
\lambda_d \left( p \right) \ge \left( \lambda'_{d-1} \left( q \right) + \left( 1-q \right) \log \left( 1-q \right) - \left( \frac{p-q}{2} \right) \log \left( \frac{p-q}{2} \right) + \right. \\
\left. \left( 1- \frac{p-q}{2} \right) \log \left( 1 - \frac{p-q}{2} \right) - \left( 1-p \right) \log \left( 1-p \right) \right). \notag
\end{align}
Since $\lambda_k \left( x \right)$ is continuous for each $x \in \left[ 0,1 \right]$ and each $k \in \mathbb{N}$ \cite{HammExist}, this will imply \eqref{lambdalower}.

\begin{lemma}
\label{setsubset}
Let $\mathcal{U}$ be the set of all subsets of $\left[ \mathbf{m'} \right]$ of cardinality $L - \ell'$.  Then
\begin{align}
\# C \left( \left[ \mathbf{m'} \right] , \ell' \right) = f_0 \sum_{U \in \mathcal{U}} \# C \left( \left[ \mathbf{m'} \right] \backslash U , \ell' \right), \quad \# \mathcal{U} = \left( \begin{array}{c} N' \\ L - \ell' \end{array} \right) \\
\mathrm{where \ } N' = \# \left[ \mathbf{m'} \right], \quad f_0 = \left( \begin{array}{c} N' - 2 \ell' \\ L - \ell' \end{array} \right)^{-1}. \notag
\end{align}
\end{lemma}

\textbf{Proof.}  Take a tiling of $\left[ \mathbf{m'} \right]$ by $\ell'$ dimers.  The monomers of this tiling are located on the subset $V$ of $\left[ \mathbf{m'} \right]$.  Clearly $\# V = N' - 2 \ell'$.  This tiling appears exactly in all $C \left( \left[ \mathbf{m'} \right] \backslash U, \ell' \right)$, where $U \subset V$.  The number of choices of these $U$ is $\left( \begin{array}{c} N' - 2 \ell' \\ L - \ell' \end{array} \right)$.  The formula for $\# \mathcal{U}$ is clear.  We have implicitly used the fact that $L - \ell' < N' - 2 \ell'$. \qed

We now consider the $d$-dimensional box $\left[ \mathbf{m} \right]$, where $\mathbf{m} = \left( \mathbf{m'}, n \right)$ and $n \ge 2$.  So $\left[ \mathbf{m} \right]$ is viewed to have $n$ horizontal levels consisting of $\left[ \mathbf{m'} \right]$.  Let $\mathbf{h}_n = \left( \ell', \dots , \ell' \right) \in \mathbb{N}^n, \mathbf{v}_n = \left( L - \ell' , \dots , L - \ell' \right) \in \mathbb{N}^{n-1}$.  For each $U \in \mathcal{U}$ we denote by $C\left( \left[ \mathbf{m} \right] , \mathbf{h}_n , \mathbf{v}_n , U \right)$ a subset of tilings in $C \left( \left[ \mathbf{m} \right] ,\mathbf{h}_n , \mathbf{v}_n \right) $ such that no dimer (vertical or horizontal) covers a spot in $U$ on the first level $\left( \left[ \mathbf{m'} \right] , 1 \right)$.

\begin{lemma}
\label{mlevels}
Let $n \ge 2$, $\mathbf{m} = \left( \mathbf{m'} , n \right)$, $\mathbf{h}_n = \left( \ell' , \dots , \ell' \right) \in \mathbb{N}^n$, $\mathbf{v}_n = \left( L-\ell' , \dots , L-\ell' \right) \in \mathbb{N}^{n-1}$.  Then
\begin{align}
\# C \left( \left[ \mathbf{m} \right], \mathbf{h}_n , \mathbf{v}_n \right) = f \sum_{U \in \mathcal{U}} \# C \left( \left[ \mathbf{m} \right], \mathbf{h}_n , \mathbf{v}_n , U \right), \quad \mathrm{where \ } f = \left( \begin{array}{c} N' - \left( L + \ell' \right) \\ L - \ell' \end{array} \right)^{-1}
\end{align}
\end{lemma}

\textbf{Proof}.  Take a tiling $\phi \in C \left( \left[ \mathbf{m} \right], \mathbf{h}_n , \mathbf{v}_n \right) $.  The monomers of this tiling, located on the first level $\left( \left[ \mathbf{m'} \right],1 \right)$, form a subset $W$ of $\left[ \mathbf{m'} \right]$.  Clearly $\#W = N' - \left( L + \ell' \right)$.  This tiling appears exactly in all $C \left( \left[ \mathbf{m} \right] ,\mathbf{h}_n , \mathbf{v}_n , U \right)$, where $U \subset W$.  The number of choices of these $U$ is $\left( \begin{array}{c} N' - \left( L + \ell' \right) \\ L - \ell' \end{array} \right)$. \qed

\begin{lemma}
\label{sumsquare}
For $n \in \mathbb{N}$
\begin{align}
\# C \left( \left[ \left( \mathbf{m'}, 2n \right) \right], \mathbf{h}_{2n} , \mathbf{v}_{2n} \right) = \sum_{U \in \mathcal{U}} \# C \left( \left[ \left( \mathbf{m'}, n \right) \right], \mathbf{h}_n , \mathbf{v}_n,U \right)^2,
\end{align}
where $C \left( \left[ \left( \mathbf{m'},1 \right) \right], \mathbf{h}_1 , \mathbf{v}_1, U \right) := C \left( \left[ \mathbf{m'} \right] \backslash U, \ell' \right)$.
\end{lemma}

\textbf{Proof}.  Observe that any tiling in $\phi \in C \left( \left[ \left( \mathbf{m'},2n \right) \right], \mathbf{h}_{2n} , \mathbf{v}_{2n} \right)$ can be obtained from exactly two tilings of $C \left( \left[ \left( \mathbf{m'},n \right) \right] , \mathbf{h}_n , \mathbf{v}_n, U \right)$, where $U$ is the location of $L - \ell'$ vertical tiles from level $n$ to level $n+1$ in $\phi$. \qed

Using Lemmas \ref{setsubset} -- \ref{sumsquare}, \eqref{mprimelambda} and the Cauchy-Schwarz inequality we obtain:

\begin{corollary}
\begin{align}
\# C \left( \left[ \left( \mathbf{m'},2 \right) \right], \mathbf{h}_2, \mathbf{v}_2 \right) & \ge \frac{e^{2N' \lambda'_{d-1} \left( q \right)}}{M f_0^2}, \quad M = \# \mathcal{U} = \left( \begin{array}{c} N' \\ L - \ell' \end{array} \right), \\
\# C \left( \left[ \left( \mathbf{m'} ,2n \right) \right], \mathbf{h}_{2n}, \mathbf{v}_{2n} \right) & \ge \frac{\left( \# C \left( \left[ \left( \mathbf{m'}, n \right) \right] , \mathbf{h}_n, \mathbf{v}_n \right) \right)^2}{Mf^2}, \quad \mathrm{for \ } n \ge 2.
\end{align}
\end{corollary}

Let $s \in \mathbb{N}$ and consider $C \left( \left[ \left( \mathbf{m'} , 2^s \right) \right], \mathbf{h}_{2^s}, \mathbf{v}_{2^s} \right)$. Use the above corollary to deduce
\begin{align}
\label{deduceineq}
\# C \left( \left[ \left( \mathbf{m'}, 2^s \right) \right], \mathbf{h}_{2^s}, \mathbf{v}_{2^s} \right) \ge Mf^2 \left( \frac{e^{N' \lambda'_{d-1} \left( q \right)}}{Mff_0} \right)^{2^s}.
\end{align}
Note that each configuration in $C \left( \left[ \left( \mathbf{m'}, 2^s \right) \right], \mathbf{h}_{2^s}, \mathbf{v}_{2^s} \right)$ has $2^s L - \left( L - \ell' \right)$ dimers.  Let $p_s = \frac{2 \left( 2^s L - \left( L - \ell' \right) \right)}{2^sN'}$.  So $\lim_{s \to \infty} p_s = p$.  \eqref{lambdalogineq} yields that
\begin{align}
\lambda_d \left( p_s \right) \ge \frac{\log \# C \left( \left[ \left( \mathbf{m'}, 2^s \right) \right], \mathbf{h}_{2^s}, \mathbf{v}_{2^s} \right)}{N' 2^s}.
\end{align}
Combine the above inequality with \eqref{deduceineq} and let $s \to \infty$ to deduce
\begin{align}
\lambda_d \left( p \right) \ge \lambda'_{d-1} \left( q \right) - \frac{\log Mff_0}{N'}.
\end{align}
Let $N' \to \infty$, i.e. $\mathbf{m'} \to \infty$, while keeping the values of $q$ and $p$ fixed, to deduce \eqref{pqchoiceineq}.

\section{An Upper Estimate of $\lambda_d \left( p \right)$}
\label{Upper}
In this section we use the notation from Section \ref{Prelim}.  We will study a sequence of cubes and use Lemma \ref{seqlemma} so that with each of these cubes, say $\mathbf{m}_i$, we need consider only a single $\mathbf{h}_i$ and $\mathbf{v}_i$.  We turn to considering a single such cube, $\mathbf{m}$, suppressing the index $i$ since we spend much time with this single cube.

Recalling notation from Section \ref{Prelim},
\begin{align}
\mathbf{m} &= \left( m_1, \dots, m_d \right), \quad \mathrm{here \ all \ } m_i = n \\
\mathbf{m'} &= \left( m_1, \dots, m_{d-1} \right) \\ \notag \\
\mathbf{h} &= \left( h_1, \dots, h_n \right) \\
\mathbf{v} &= \left( v_1, \dots, v_{n-1} \right)
\end{align}
$h_i$ is the number of horizontal dimers in the $i^{\mathrm{th}}$ layer, and $v_i$ the number of vertical dimers connecting the $i^{\mathrm{th}}$ and $\left(i+1 \right)^{\mathrm{th}}$ layers.  $V_i$ is the location in $\mathbf{m'}$ of the vertical dimers connecting the $i^{\mathrm{th}}$ and $\left( i+1 \right)^{\mathrm{th}}$ layer.  We set as before
\begin{align}
N' = \# \mathbf{m'} = n^{d-1}
\end{align}
and take $\ell$ to be the total number of dimers.  We then have the following.
\begin{align}
& V_i \cap V_{i+1} = \varnothing \quad i = 1, \dots , n-2 \label{visect} \\
& v_i + v_{i+1} \le N' \quad i = 1, \dots , n-2 \label{vsum} \\ \notag \\
& 2h_i + v_{i-1} + v_i \le N' \quad i = 2, \dots , n-2 \label{hvineq} \\
& \sum h_i + \sum v_i = \ell \label{hvsum}
\end{align}
(There are slight obvious modifications of \eqref{visect} -- \eqref{hvineq} for $i = 1, n-1, n$.)

For our given $\mathbf{v}$ we count the possible choices of $V_1 , \dots , V_{n-1}$ such that $\# V_i = v_i$ for $i = 1, \dots , n-1$.  We can choose $V_1$ in $\left( \begin{array}{c} N' \\ v_1 \end{array} \right)$ ways, $V_2$ in $\left( \begin{array}{c} N' -v_1 \\ v_2 \end{array} \right)$, etc.  Thus we arrive at
\begin{align}
\alpha \left( N' , \mathbf{v} \right) = \left( \begin{array}{c} N' \\ v_1 \end{array} \right) \prod_{i=2}^{n-1} \left( \begin{array}{c} N' - v_{i-1} \\ v_i \end{array} \right)
\end{align}
for the total number of choices.  Using this enumeration, and recalling Lemma \ref{seqlemma}, we will get an upper bound for $\lambda_d \left( p \right)$ as the limit as $n \to \infty$ of the maximum possible value of
\begin{align}
\label{maxlim}
\frac{1}{n^d} \left( \log \alpha \left( n^{d-1}, \mathbf{v} \right) + \log \prod_{i=1}^n C \left( \left[ \mathbf{m'} \right] \backslash V_{i-1} \cup V_i , h_i \right) \right)
\end{align}
We define
\begin{align}
p &= \frac{2 \ell}{n^d} \\
u_i &= \frac{2 v_i}{n^{d-1}}, \quad t_i = \frac{2h_i}{n^{d-1}} \\
u &= \frac{1}{n} \sum_{i=1}^n u_i , \quad t = \frac{1}{n} \sum_{i=1}^n t_i \\
\tilde{t}_i &= \frac{2h_i}{ n^{d-1} \left( 1 - \frac{1}{2} \left( u_i + u_{i-1} \right) \right)}
\end{align}
There follows
\begin{align}
u+t=p
\end{align}
$u$ is the density of vertical dimers and $t$ the density of horizontal dimers.  $\tilde{t}_i$ is the fraction of the vertices in $\left[ \mathbf{m'} \right] \backslash \left( V_{i-1} \cup V_i \right)$ covered by dimers.

Using Conjecture \ref{estim} we may effectively assert
\begin{align}
\frac{\log C \left( \left[ \mathbf{m'} \right] \backslash \left( V_{i-1} \cup V_i \right), h_i \right)}{\left( 1 - \frac{u_{i-1} + u_i }{2} \right) n^{d-1} } \le \lambda_{d-1} \left( \tilde{t}_i \right).
\end{align}
Hence
\begin{align}
\frac{\log \prod_{i=1}^n C \left( \left[ \mathbf{m'} \right] \backslash \left( V_{i-1} \cup V_i \right), h_i \right)}{n^d} \le \sum_{i=1}^n \frac{ 1 - \frac{u_{i-1} + u_i}{2}}{n} \lambda_{d-1} \left( \tilde{t}_i \right).
\end{align}
Recall that $\lambda_{d-1} \left( p \right)$ is a concave function.  (See \cite{Friedlandetal} for a sharper statement.)  Since $\frac{1}{n} \sum_{i=1}^n \left(  1 - \frac{u_{i-1} + u_i}{2} \right) = 1-u$ it follows that
\begin{align}
\sum_{i=1}^n \frac{1 - \frac{u_{i-1} + u_i}{2}}{n} \lambda_{d-1} \left( \tilde{t}_i \right) \le \left( 1-u \right) \lambda_{d-1} \left( t' \right), \quad t' = \sum_{i=1}^n \frac{1 - \frac{u_{i-1} + u_i}{2}}{\left( 1-u \right)n} \tilde{t}_i.
\end{align}
In view of (4.15) we obtain
\begin{align}
t' = \sum_{i=1}^n \frac{1 - \frac{u_{i-1} + u_i}{2}}{\left( 1-u \right) n} \frac{2h_i}{n^{d-1} \left( 1 - \frac{u_{i-1} +u_i}{2} \right)} = \frac{1}{\left( 1-u \right) n^d} \sum_{i=1}^n 2h_i = \frac{t}{\left( 1-u \right)}.
\end{align}
We combine with the previous inequalities to deduce that
\begin{align}
\frac{\log \prod_{i=1}^n C \left( \left[ \mathbf{m'} \right] \backslash \left( V_{i-1} \cup V_i \right), h_i \right)}{n^d} \le \left( 1-u \right) \lambda_{d-1} \left( \frac{t}{\left( 1-u \right)} \right).
\end{align}

To bound the first term in \eqref{maxlim} it is straightforward to show the maximum is achieved with all  $v_i$ equal, $v_i = \frac{N' u}{2}$.  This yields
\begin{align}
\limsup_{n \to \infty} \frac{\log \alpha \left( n^d, \mathbf{v} \right)}{n^d} \le \left( 1 - \frac{u}{2} \right) \log \left( 1 - \frac{u}{2} \right) - \frac{u}{2} \log \frac{u}{2} - \left( 1-u \right) \log \left( 1-u \right).
\end{align}
In conclusion we derive the following.
\begin{align}
\label{conclusion}
\lambda_d \left( p \right) \le \max_{u} \left\{ \left( 1- \frac{u}{2} \right) \log \left( 1 - \frac{u}{2} \right) - \frac{u}{2} \log \frac{u}{2} - \left( 1-u \right) \log \left( 1-u \right) + \left( 1-u \right) \lambda_{d-1} \left( \frac{p-u}{1-u} \right) \right\}
\end{align}
Because the expected value of $u$ is $\frac{p}{d}$, by the symmetry of the cube, one can show one can restrict the range of $u$ in taking the maximum in \eqref{conclusion} to either $\left[ 0, \frac{p}{d} \right]$ or $\left[ \frac{p}{d}, 1 \right] $.  This leads to two formulas for an upper bound.  Unless these are equal, one gets a better bound restricting to one or the other of these ranges.

One can employ the argument behind Lemma \ref{setsubset} in the construction of this section, obtaining an alternate upper bound:
\begin{align}
\label{bound2}
\lambda_d \left( p \right) \le \max_u \left\{ -\frac{u}{2} \log \frac{u}{2} - \left( 1-p \right) \log \left( 1-p \right) + \left( 1-p + \frac{u}{2} \right) \log \left( 1-p+ \frac{u}{2} \right) \right. \notag \\
+ \left. \left( 1 - \frac{u}{2} \right) \lambda_{d-1} \left( \frac{p-u}{1 - \frac{u}{2}} \right) \right\}
\end{align}
Again the range of $u$ may be restricted to either $\left[ 0 , \frac{p}{d} \right]$ or $\left[ \frac{p}{d} , 1 \right]$.  We outline this alternate development, closely following the previous route, and using the same notation.

We let $W_r \left( V_r \right)$ be the number of configurations in the portion of the lattice living in layers $r+1, r+2, \dots, n$ (summing over both vertical and horizontal dimers).  This will involve summing over $V_{r+1}, \dots, V_{n-1}$ compatible with $\# \left( V_i \right) = v_i, V_i \cap V_{i+1} = \varnothing$, and $h_i$ horizontal dimers in layer $i$ for $i=r+1, \dots, n$.  $V_r, h_{r+1}, \dots, h_n, v_{r+1}, \dots, v_{n-1}$ are fixed.  Then we have
\begin{align}
W_{n-1} \left( V_{n-1} \right) & \le C \left( \left[ m' \right] \backslash V_{n-1}, h_n \right) \notag \\
& \le e^{n^{d-1}\left( 1 - u_{n-1}/2 \right) \lambda_{d-1} \left( \tilde{t}_n \right)} \equiv b_{n-1}
\end{align}
with
\begin{align}
\tilde{t}_i = \frac{2h_i}{n^{d-1} \left( 1 - u_{i-1}/2 \right)}
\end{align}

We find the bounds $b_i$ inductively.  Assume we have bounds $b_{r+1}, b_{r+2}, \dots , b_{n-1}$.  We seek $b_r$.
\begin{align}
W_r \left( V_r \right) \le \sum_{ \begin{array}{c}  V_{r+1} \\ \# \left( V_{r+1} \right) = v_{r+1} \\ V_{r+1} \cap V_r = \varnothing \end{array} } C \left( \left[ m' \right] \backslash V_r \cup V_{r+1}, h_{r+1} \right) b_{r+1}
\end{align}
By the argument of Lemma \ref{setsubset}
\begin{align}
& \le C \left( \left[ m' \right] \backslash V_r, h_{r+1} \right) b_{r+1} \left( \begin{array}{c} N' - 2 h_{r+1} - v_r \\ v_{r+1} \end{array} \right) \\
& \le e^{n^{d-1} \left( 1 - u_r/2 \right) \lambda_{d-1} \left( \tilde{t}_{r+1} \right)} b_{r+1} \left( \begin{array}{c} N - 2 h_{r+1} - v_r \\ v_{r+1} \end{array} \right)
\end{align}
As before
\begin{align}
\lambda_d \left( p \right) & \le \mathrm{I} \cdot \mathrm{I\kern -0.2ex I} \notag \\
\mathrm{I} & = \sum \frac{1}{n} \left( 1 -  \frac{u_i}{2} \right) \lambda_{d-1} \left( \tilde{t}_{i+1} \right)
\end{align}
which by a counting argument
\begin{align}
\le \frac{t}{1-\frac{u}{2}}
\end{align}
and
\begin{align}
\mathrm{I\kern -0.2ex I} = \frac{1}{n^d} \sum \log \left( \begin{array}{c} N' - 2 h_{r+1} - v_r \\ v_{r+1} \end{array} \right)
\end{align}
The maximum in $\mathrm{I\kern -0.2ex I}$ occurs where all $v_i$ are equal and all $h_i$ equal (at least for $i=2, \dots , n-2$, just as good).  This leads to \eqref{bound2}.

We now study the relation between \eqref{upperbound} and \eqref{betterbound} (or \eqref{conclusion} and \eqref{bound2}).  We take $\lambda_d \left( p \right)$ to be given by its approximate form
\begin{align}
\label{lambdaform}
\tilde{\lambda}_d \left( p \right) = \frac{1}{2} p \log \left( 2d \right) - \frac{p}{2} \log p - \left( 1-p \right) \log \left( 1-p \right) - \frac{p}{2}
\end{align}
and substitute this into
\begin{align}
&\left\{ \left( 1 - \frac{u}{2} \right) \log \left( 1 - \frac{u}{2} \right) - \frac{u}{2} \log \left( \frac{u}{2} \right) - \left( 1 - u \right) \log \left( 1 - u \right) + \left( 1 - u \right) \lambda_{d-1} \left( \frac{p-u}{1-u} \right) \right\} - \notag \\
&\left\{ \left( 1 - p + \frac{u}{2} \right) \log \left( 1 - p + \frac{u}{2} \right) - \frac{u}{2} \log \left( \frac{u}{2} \right) - \left( 1 - p \right) \log \left( 1 - p \right) + \left( 1 - \frac{u}{2} \right) \lambda_{d-1} \left( \frac{p-u}{1 - \frac{u}{2}} \right) \right\}
\end{align}
Surprisingly, by a simple computation this comes out to be
\begin{align}
\frac{1}{2} \left( p - u \right) \log \left( \frac{1 - \frac{u}{2} }{1 - u} \right)
\end{align}
In the region $u<p$ this is positive.  This shows that the expression in braces in \eqref{upperbound} is pointwise (in $u$) greater than the expression in braces in \eqref{betterbound}.  Thus \eqref{betterbound} is a better upper bound than \eqref{upperbound} for $\lambda_d \left( p \right)$ as given by \eqref{lambdaform}.

For $0 \le u \le p/d$ and $d \ge 2$ we have
\begin{align}
\label{logineq}
\frac{1}{2} \left( p - u \right) \log \left( \frac{1 - \frac{u}{2}}{1 - u} \right) > c_1 p u
\end{align}
We let the actual $\lambda_d \left( p \right)$ be related to the approximate form, \eqref{lambdaform}, as
\begin{align}
\label{actuallambda}
\lambda_d \left( p \right) = \tilde{\lambda}_d \left( p \right) + E_d \left( p \right)
\end{align}
We assume for $d > d_0$ one has
\begin{align}
\label{assume1}
\left| E_d \left( p \right) \right| \le c_2 \frac{p}{d}
\end{align}
and
\begin{align}
\label{assume2}
\left| \frac{d}{dp} E_d \left( p \right) \right| \le c_3 \frac{1}{d}
\end{align}
Then there is a $\bar{d}$ such that if $d > \bar{d}$ then \eqref{betterbound} provides a better upper bound than \eqref{upperbound}.  That is if one substitutes (for $d > \bar{d}$) \eqref{actuallambda} into the right side of \eqref{upperbound} one gets a larger value than if one substitutes into the right side of \eqref{betterbound}.  One has used the bound
\begin{align}
\label{absbound}
\left| \left( 1 - u \right) E_{d-1} \left( \frac{p-u}{1-u} \right) - \left( 1 - \frac{u}{2} \right) E_{d-1} \left( \frac{p-u}{1-\frac{u}{2}} \right) \right| \le & \\
c_4 u \left| E_{d-1} \left( p^* \right) \right| + c_5 p u \left| E'_{d-1} \left( p^{**} \right) \right| &
\end{align}
Provided \eqref{assume1} and \eqref{assume2} are satisfied by $E_d \left( p \right)$ it is easy to deduce, using \eqref{logineq} and \eqref{absbound}, that \eqref{betterbound} is a better bound than \eqref{upperbound} for $d$ large enough.  One looks at the $u^*$ that maximizes the expression in braces in \eqref{betterbound}.  By \eqref{logineq} and \eqref{absbound}, for large enough $d$, the expression in braces in \eqref{upperbound} evaluated at $u^*$ is larger than the expression in braces in \eqref{betterbound} at $u^*$.  Thus the bound \eqref{betterbound} beats the bound \eqref{upperbound}.

\section{The Asymptotic Expansion}
\label{Asymptotic}
This section largely parallels the development in \cite{FedTerm}, with modifications appropriate to the monomer-dimer problem.  We do not assume familiarity with \cite{FedTerm}.

We work with a periodic cubical lattice, $\Lambda$, of even edge length.  We denote the volume, or equivalently the number of vertices, by either $V$ or $N$.  A fraction $p$ of the vertices are covered by dimers, thus $pN/2$ dimers are used in each covering.  We will be using the terms `$p$-tiling' and `$p$-cover' to describe coverings of a fraction $p$ of the vertices.

In $d$ dimensions there are $d$ `kinds' of dimers, each kind oriented in one of the $d$ lattice directions.  Each kind of dimer may be `located' at some place on the lattice.  We generalize this situation as follows.  A `located tile' is a two element subset of the lattice.  A `tile' is an equivalence class of located tiles, with equivalence given by setting two subsets equivalent if one is a translation of the other.  (A `tile' is a generalization of a `kind of dimer'.)  Notice that tiles need not be connected.

We consider tiles with a `weighting', a function on tiles.  We normalize the weightings we consider, by requiring, if $g$ is the weighting function,
\begin{align}
\label{weightsum}
\sum_t g \left( t \right) = 1/2,
\end{align}
the sum over all tiles, $t$.  We let $f$ be the weighting function given by
\begin{align}
f \left( t \right) = \left\{ \begin{array}{l}
\frac{1}{2d} {\rm \ if \ } t {\rm \ is \  a \  dimer,} \\[.05in]
0 {\rm \ otherwise,}
\end{array} \right.
\end{align}
that clearly satisfies the normalization condition (\ref{weightsum}).

A `$p$-tiling' $T_i$ of $\Lambda$ is a set of two element subsets of $\Lambda$,
\begin{align}
T_i = \left\{ s_1^i, s_2^i, \dots , s_{pN/2}^i \right\},
\end{align}
where the $s_k^i$ are disjoint, they are located tiles.

We now realize the sum over all possible dimer $p$-covers of $\Lambda$, the goal of our study, as
\begin{align}
\left( 2d \right)^{pN/2}Z
\end{align}
with
\begin{align}
Z = \sum_{T_i} \prod_{s_{\alpha} \in T_i} f \left( \bar s_{\alpha} \right),
\end{align}
where the sum is over all $p$-tilings of $\Lambda$; the product over the $f$'s selects those tilings in which all the tiles employed are dimers.  The bar over a subset indicates the equivalence class of the subset as defined previously.

We let $f_0$ be a constant function on tiles with value $\frac{1}{N-1}$, to satisfy the normalization condition (\ref{weightsum}).  We write
\begin{align}
f = f_0 + \left( f - f_0 \right) \equiv f_0 + v,
\end{align}
$f$, $f_0$, and $v$ all functions on tiles.  $Z$ becomes
\begin{align}
Z &= \sum_{T_i} \prod_{s_{\alpha} \in T_i } \left( f_0 + v \left( \bar s_{\alpha} \right) \right), \label{Zsumprod} \\
Z &= Z_0 + Z_1 + Z_2 + \cdots \label{Zsum}
\end{align}
having expanded $Z$ in powers of $v$.

We note
\begin{align}
e^{N \lambda_d \left( p \right)} = \left( 2d \right)^{Np/2} Z.
\end{align}
Here $\lambda_d \left( p \right)$ is understood to be a function of $N$, the usual $\lambda_d \left( p \right)$ the infinite volume limit
\begin{align}
\label{lambdad-Z}
\lambda_d \left( p \right) = \frac{p}{2} \log \left( 2d \right) + \frac{1}{N} \log Z.
\end{align}
If one replaces $Z$ by $Z_0$, the mean field approximation in a natural nomenclature, one gets taking the infinite volume limit
\begin{align}
\label{lambdapprox}
\lambda_d \left( p \right) & \cong \frac{p}{2} \log \left( 2d \right) - \frac{p}{2} \log p - \left( 1 - p \right) \log \left( 1 - p \right) - \frac{p}{2} \notag \\
&= \frac{p}{2} \log \left( 2d \right) + \lim_{N \to \infty} \frac{1}{N} \log Z_0
\end{align}
by an easy calculation.

We return to (\ref{Zsum}) and introduce some convenient notations:
\begin{align}
Z &= Z_0 Z^*, \label{ZZstar} \\
Z^* &= 1 + Z^*_1 + Z^*_2 + \cdots , \label{Zstarsum} \\
Z^*_i &= Z_i / Z_0.
\end{align}
There is a natural factorization of $Z_i$ into a contribution from the factors of $v$ in (\ref{Zsumprod}) which we call $\bar Z_i^*$ and the factors of $f_0$ in (\ref{Zsumprod}) which we call $\beta \left( N,i \right) Z_0$ so that
\begin{align}
\label{factorbeta}
Z_i^* = \beta \left( N,i \right) \bar Z_i^*
\end{align}
with
\begin{align}
\label{betapprox}
\beta \left( N, jN \right) \sim e^{N H \left( p,j \right)}
\end{align}
where
\begin{align}
\label{Hpj}
H \left( p,j \right) = \left( 1 - 2j \right) \log \left( 1 - 2j \right) + j + \frac{p}{2} \log p - \left( \frac{p}{2} - j \right) \log \left( p - 2j \right).
\end{align}
Equations (\ref{betapprox}) and (\ref{Hpj}) follow from a short computation, always working in the large $N$ limit.

We let $\tilde{Z}^*$ be $Z^*$ with $\beta \left( N,i \right)$ replaced by 1,
\begin{align}
\label{Ztildesum}
\tilde{Z}^* = 1 + \bar Z_1^* + \bar Z_2^* + \cdots,
\end{align}
where a detailed specification of $\bar Z^*_i$ is given by
\begin{align}
\bar Z_i^* = \frac{1}{i!} \sum_{\begin{array}{c} s_1, s_2, \dots , s_i \\ {\rm \ disjoint} \end{array} } \prod_{\alpha = 1}^i v \left( \bar s_{\alpha} \right).
\end{align}
Now referring to \cite{Brydges} we may write a cluster expansion for $\tilde{Z}^*$
\begin{align}
\log \tilde{Z}^* &= \sum_s \frac{1}{s!} J_s, \label{clustexpln} \\
J_s &= \sum_{s_1, s_2 , \dots , s_s} v \left( \bar s_1 \right) \cdots v \left( \bar s_s \right) \psi'_c \left( s_1, s_2, \dots , s_s \right) \label{Js} ,
\end{align}
where we may identify Eq. (2.5a) of \cite{Brydges} with $\tilde{Z}^*$, and (\ref{clustexpln}), (\ref{Js}) with Eq. (2.7) of \cite{Brydges}.  The located tiles in the sum of (\ref{Js}) are forced to overlap so they cannot be divided into two disjoint sets.  $\psi'_c$ is a numerical factor depending on the overlap pattern.  To make our computations mathematically rigorous it will be necessary to study the convergence properties of sums such as in (\ref{clustexpln}).  At present this appears very difficult, and we by no means see yet a clear route to a proof.  It is a challenging problem for the mathematical physicist.

It is easy to show $J_1 = 0$ and it is proven in \cite{FedDim} that
\begin{align}
\label{expandJs}
J_s = \frac{C_{s,r}}{d^r} + \frac{C_{s,r+1}}{d^{r+1}} + \cdots + \frac{C_{s,s-1}}{d^{s-1}}
\end{align}
with $r \ge s/2$.  We also find it convenient to define
\begin{align}
N \bar J_i = \left( 1/i! \right) J_i.
\end{align}

From (\ref{Zstarsum}), (\ref{factorbeta}), (\ref{Ztildesum}), (\ref{clustexpln}) one gets
\begin{align}
\label{Zstarbigsum}
Z^* = \sum_{\alpha_1 , \dots , \alpha_{s+1}} \beta \left( N , \sum i \alpha_i \right) \bar J_1^{\alpha_1} \cdots \bar J_{s+1}^{\alpha_{s+1}} \frac{N^{\sum \alpha_i}}{\alpha_1 ! \cdots \alpha_{s+1} !}.
\end{align}
We approximate the sum in (\ref{Zstarbigsum}) by its largest term, in the limit $N \to \infty$.  If all the $\bar J$'s are positive this is a reasonable way to extract the dominant asymptotic limit.  In \cite{FedCon} an argument is given that our results will hold even if some of the $\bar J$'s are negative.

Most important to observe is that the $J$'s of this paper are the same as the $J$'s of the dimer problem!  The $J_i$ were computed through $J_6$ as follows, see \cite{FedTerm}:
\begin{align}
\bar J_1 &= 0, \label{J1} \\
\bar J_2 &= \frac{1}{8} \frac{1}{d}, \label{J2} \\
\bar J_3 &= \frac{1}{12} \frac{1}{d^2}, \label{J3} \\
\bar J_4 &= - \frac{3}{32} \frac{1}{d^2} + \frac{3}{64} \frac{1}{d^3}, \label{J4} \\
\bar J_5 &= - \frac{1}{8} \frac{1}{d^3} - \frac{3}{80} \frac{1}{d^4}, \label{J5} \\
\bar J_6 &= \frac{7}{48} \frac{1}{d^3} - \frac{5}{64} \frac{1}{d^4} - \frac{1}{6} \frac{1}{d^5}. \label{J6}
\end{align}

The computations were done in integral arithmetic using Maple.  Because of (\ref{expandJs}), to compute the $J_i$ through $J_6$ for all $d$, it is sufficient to compute these $J_i$ for one, two, and three dimensions.  The most complicated computation of these involved placing down six dimers in three dimensions, in all possible ways where overlaps make it impossible to disconnect into two disjoint subsets of dimers.  One is using an exponential time algorithm, $ J_5$'s computation took 3 seconds and $J_6$'s two weeks.  To compute the $1/d^4$ terms in (\ref{possexp}) would require knowledge of $J_7$ and $J_8$ also, we think unless a new method is found one will never do this computation!

We return to computing the largest term in (\ref{Zstarbigsum}).  We differentiate with respect to the $\alpha$'s in (\ref{Zstarbigsum}), using (\ref{betapprox}) and (\ref{Hpj}) to deal with the factor of $\beta$.  Finally scaling $\alpha_i \to \frac{1}{N} \alpha_i$ we find
\begin{align}
\log \alpha_k = \log \bar J_k + \frac{\partial}{\partial \alpha_k} H \left( p , \sum i \alpha_i \right)=
\log \bar J_k +F_k
\end{align}
and
\begin{align}
Z^* \sim e^{N \left\{ - \sum \alpha_i F_i + \sum \bar J_i e^{F_i} + H \left( p, \sum i \alpha_i \right) \right\} }.
\end{align}
From these two equations, and (\ref{lambdad-Z}), (\ref{lambdapprox}), (\ref{ZZstar}), one can find $\lambda_d \left( p \right)$ as a formal power series in the $J$'s, and then a formal power series in $1/d$ using the expressions (\ref{J1}) through (\ref{J6}) for the $J$'s.  This algebra we did using a 40-line Maple program, resulting in \eqref{lambdaseries} -- \eqref{c3}.

In \cite{Fed11} the first named author shows that if the terms in the formal expansion \eqref{lambdaseries} are rearranged as a power series in $p$, then for sufficiently small $p$ this series converges.

\section{The LAMC and Other Conclusions}
\label{Conclusions}
For any infinite graph $G=(V,E)$ on a countable number of vertices $V$,
where each vertex $V$ has at most degree $r$, one can define the notion of the $p$-entropy $\lambda_G(p)$,  similarly to the definition of the entropy $\lambda_{\mathcal{U}}(p)$ defined in \S2.  The \emph{Lower Asymptotic Matching Conjecture} for infinite regular $r$-bipartite graphs
claims that $\lambda_G(p)$ is bounded below by a universal function $\omega_r(p)$, see equations (1.3) and (1.4) of \cite{Friedlandetal}.
$\omega_r(p)$ can be considered as the $p$-entropy of a \emph{random} $r$-regular infinite bipartite graph.  The formula for $\omega_{2d}(p)$ is given by the right-hand side of \eqref{LAMC}.
We first prove the LAMC for a rectangular lattice, 
\begin{theorem}
The LAMC is true for a rectangular lattice.  That is
\begin{align}
\label{LAMC}
\lambda_d \left( p \right) \ge \frac{1}{2} \left[ p \log \left( 2d \right) - p \log p - 2 \left( 1-p \right) \log \left( 1-p \right) + \left( 2d-p \right) \log \left( 1 - \frac{p}{2d} \right) \right]
\end{align}
\end{theorem}
\textbf{Proof.}  We prove this by induction.  It is true for $d=1$ by \cite{Friedlandetal} eq (1.1) or \cite{FriedPeledComp} $\S$ 4.  In the inductive step we are given
\begin{align}
\lambda_{d-1} \left( p \right) \ge \frac{1}{2} \left[ p \log \left( 2 \left( d-1 \right) \right) - p \log p - 2 \left( 1-p \right) \log \left( 1-p \right) + \left( 2 \left( d-1 \right) - p \right) \log \left( 1 - \frac{p}{2 \left( d-1 \right)} \right) \right]
\end{align}
We substitute this into (the right side of) \eqref{recur}, proven in Theorem \ref{mainestimate}, and arrive by a simple computation at exactly \eqref{LAMC}.

We turn to a study of the relation between our upper and lower recursive bounds and the asymptotic expansion.  We assume that $\lambda_d \left( p \right)$ is given by an asymptotic expansion in inverse powers of $d$, whose first few terms are as in \eqref{ansatz}.
\begin{align}
\label{asympexp}
\lambda_d \left( p \right) \sim \frac{1}{2} p \log \left( 2d \right) - \frac{p}{2} \log \left( p \right) - \left( 1-p \right) \log \left( 1-p \right) - \frac{p}{2} + \frac{ap^2}{d} + \frac{bp^4 +cp^3}{96d^2} + \dots
\end{align}

\begin{theorem}
\label{parts}
\hfill
\begin{enumerate}[A)]
\item $a \ge \frac{1}{8}$
\item If $a = \frac{1}{8}$, then $\left( - 2 + 2bp + c \right) \ge 0$
\item If $a = \frac{1}{8}$ and $c = 2$, then $b \ge 0$
\item Then values of $a,b,c$ as given in \eqref{c1} and \eqref{c2} ensure that the asymptotic expansion in \eqref{lambdaseries} satisfies the implications of the lower recursive inequality \eqref{recur}.
\end{enumerate}
\end{theorem}
\textbf{Proof.}  If we substitute \eqref{asympexp} into the recursive inequality \eqref{recur}, then the value of the right side minus the left side is given in \eqref{rightminusleft} which we rewrite
\begin{align}
\label{r-lrewrite}
- \frac{1}{8} \left( -1 + 8a \right) \frac{p^2}{d^2} - \frac{1}{96} \left( -2 + 2bp + c \right) \frac{p^3}{d^3} + \mathcal{O} \left( \frac{1}{d^4} \right)
\end{align}
The requirement that \eqref{r-lrewrite} is asymptotically $\le 0$ yields parts \textit{A, B,} and \textit{C} of the theorem.  The values of $a,b,c$ as given in \eqref{c1} and \eqref{c2} ensure \eqref{r-lrewrite} is asymptotically negative.

In short our asymptotic expansion for $\lambda_d \left( p \right)$, \eqref{lambdaseries}, satisfies the lower recursive inequality.

The upper recursive inequalities we have \eqref{upperbound}, \eqref{betterbound} are not strong enough to put conditions on the coefficients of the asymptotic series, such as parts \textit{A,B,} and \textit{C} of Theorem \ref{parts}. 

\section{Numerical tests for $d=1,2,3$}
\label{numtests}
In the last section we saw our expansion satisfied the implications of the lower recursive bound asymptotically.  Here we study this expansion and our bounds for small $d$, $d=1,2,3$.

We find it convenient to rearrange our expansion for $\lambda_d (p)$ as a power series in $p$, perhaps this is always a better form to work with.
\begin{align}
\label{ArrangePowerSeries}
\lambda_d (p) \sim \frac{1}{2} \left( p \ln (2d) - p \ln p - 2 (1-p) \ln (1-p) - p \right) + \sum_{k=2} a_k(d) p^k
\end{align}
where we see from \eqref{lambdaseries} -- \eqref{c3} that 
\begin{align}
a_2 (d) &= \frac{1}{8} \frac{1}{d} \\
a_3 (d) &= \frac{1}{48} \frac{1}{d^2} \\ 
a_4 (d) &= \frac{1}{32} \frac{1}{d^2} - \frac{5}{192} \frac{1}{d^3}.
\end{align}
using the fact mentioned after \eqref{lambdaseries} that $c_k \left( p \right)$ is built up of powers $p^s$ with $k < s \le 2k$.  Knowing $\bar{J}_i$ for $i \le 6$ we can also compute:
\begin{align}
a_5 (d) = \frac{1}{16} \frac{1}{d^3} - \frac{39}{640} \frac{1}{d^4} \\
a_6 (d) = \frac{1}{24} \frac{1}{d^3} - \frac{1}{32} \frac{1}{d^4} - \frac{19}{1920} \frac{1}{d^5}
\end{align}
 
\subsection{$d=1$}

Recall the exact formula for $\lambda_1(p)$, e.g. \cite[end of \S4]{FriedPeledComp}
\begin{align}
\label{exactlambda}
\lambda_1(p)=
 \left( 1 - \frac{p}{2} \right) \log \left(1 - \frac{p}{2} \right) - \frac {p}{2} \log\left(
 \frac{p}{2}\right) - (1 - p)\log (1 - p).
\end{align}
Expand $\log \left(1 - \frac{p}{2}\right)$ in power series in $p$ to deduce that
\begin{align}
\label{expandlog}
&&\lambda_1 \left( p \right) = \frac{1}{2} p \log \left( 2 \right) - \frac{p}{2} \log \left( p \right) - \left( 1-p \right) \log \left( 1-p \right) - \frac{p}{2}+
\sum_{k=2}^{\infty}\frac{p^{k}}{(k-1)k2^{k}}.
\end{align}

It is easy to check the values of $a_k (1)$ are correctly given to match \eqref{expandlog} for $k=2, \dots , 6$.  (We have not tried to prove this to all orders, the $\bar{J}_i$ for $d=1$ can all be computed, but it probably can be done.)  This is a remarkable validation of our procedure to compute $\lambda_d (p)$.

\subsection{The Expansion for $\lambda_2 (1)$ and $\lambda_3 (1)$}
In eq. \eqref{ArrangePowerSeries} we consider putting the upper limit in the sum to be $n$.  Then for $n=2, \dots , 6$ we get a sequence of ``approximations'' to $\lambda_d (p)$.  Setting $d=2$, $p=1$ we get the sequence
\begin{align}
\label{approxd2p1}
.2556, .2609, .2654, .2694, .2724
\end{align}
and setting $d=3$, $p=1$ we get
\begin{align}
\label{approxd3p1}
.4375, .4399, .4424, .4439, .4450.
\end{align}
The exact value for $\lambda_2 (1)$ is given as
\begin{align}
\lambda_2 (1) = .2915 \dots
\end{align}
by \cite{Fisher} and \cite{Kasteleyn}, and $\lambda_3 (1)$ satisfies
\begin{align} 
.440075 \le \lambda_3 (1) \le .457547
\end{align}
by \cite{Ciucu} and \cite{Friedlandetal}.  We may well believe the sequence \eqref{approxd2p1} extended converges to $\lambda_2 (1)$ and \eqref{approxd3p1} extended converges to $\lambda_3 (1)$.  We now believe \eqref{ArrangePowerSeries} converges for all physical $p$, not just small $p$ as proved in \cite{Fed11}!

\subsection{The Expansion and Bounds for $\lambda_2 (p)$}
We present a table of values for $\lambda_2 (p)$, its bounds and approximations.  The first column gives the value of $p$.  The second column, the expansion value, is the result of keeping in \eqref{ArrangePowerSeries} the $a_k (2)$ from $k=2$ to $k=6$, the known values.  The third column is the rigorous lower bound \eqref{recur}.  The fourth column presents the ``exact'' value for $\lambda_2 (p)$, actually a very good approximation from \cite{FriedPeledComp} and \cite{Baxter}.  Looking at the table on page 654 of Baxter's paper \cite{Baxter}, a row with the three column entries $c1, c2, c3$ corresponds to $p=2c_3$, $\lambda_2 (p) = \ln \left( c_2 \right) - (1-p) \ln \left( c_1 \right)$.  The values of $p$ thus in this table led to the choices for $p$ in our table.  The final column is the upper bound from \eqref{betterbound}, actually rigorous for $d=2$.
\begin{center}
\begin{tabular}{l | l | l | l | l}
p & exp & lb & exact & ub \\
\hline 
0 & 0 & 0 & 0 & 0 \\
.14870 & .30887 & .30887 & .30887 & .31030 \\
.26030 & .45283 & .45281 & .45284 & .45734 \\
.50426 & .63492 & .63449 & .63495 & .65274 \\
.77053 & .62983 & .62678 & .63086 & .67319 \\
1 & .27236 & .26162 & .29156 & .34657
\end{tabular}
\end{center}
We are struck by how much better the lower bound is than the upper bound, and how good an approximation is the expansion, keeping the terms we know, for $p \lesssim .5$.

\section{Future Directions}
We leave to future research three compelling problems:  prove that the ``asymptotic'' expansion expressed as a power series in $p$ converges for all physical $p$ and $d$, find better upper bounds pure or recursive, prove Conjecture \ref{estim}.

\end{document}